\documentclass[prb,superscriptaddress,twocolumn]{revtex4-2}
\usepackage{amsmath}
\usepackage{amssymb}
\usepackage{bm}
\usepackage{epsfig}
\usepackage{graphicx}
\usepackage{color}
\usepackage[unicode=true,colorlinks=true,citecolor=blue,urlcolor=blue]{hyperref} 
\usepackage[utf8]{inputenc}
\usepackage[english]{babel}
\usepackage{xcolor}
\usepackage{gensymb}
\usepackage{physics}
\usepackage{placeins}

\usepackage{tikz}
\usepackage[compat=1.1.0]{tikz-feynman}

\usepackage[normalem]{ulem}

\begin{document}
\title{Spin relaxation of localized electrons in monolayer MoSe$_2$:\\ importance of random effective magnetic fields
}
\author{Ey\"up Yalcin}
\author{Ina V. Kalitukha}
\author{Ilya A. Akimov}
\author{Vladimir L. Korenev}
\author{Olga S. Ken}
\affiliation{Experimentelle Physik 2, Technische Universit\"at Dortmund, 44221 Dortmund, Germany}
\author{Jorge Puebla}
\affiliation{RIKEN Center for Emergent Matter Science, Saitama 351-0198, Japan}
\author{Yoshichika Otani}
\affiliation{RIKEN Center for Emergent Matter Science, Saitama 351-0198, Japan}
\affiliation{Institute for Solid State Physics, University of Tokyo, Kashiwa 277-8581, Japan} 
\author{Oscar~ M.~Hutchings}
\author{Daniel J. Gillard}
\author{Alexander I. Tartakovskii}
\affiliation{Department of Physics and Astronomy, The University of Sheffield, Sheffield S3 7RH, UK}
\author{Manfred Bayer}
\affiliation{Experimentelle Physik 2, Technische Universit\"at Dortmund, 44221 Dortmund, Germany}

\begin{abstract}
We study the Hanle and spin polarization recovery effects on resident electrons in a monolayer MoSe$_2$ on EuS. We demonstrate that localized electrons provide the main contribution to the spin dynamics signal at low temperatures below 15~K for small magnetic fields of only a few mT. The spin relaxation of these electrons is determined by random effective magnetic fields due to a contact spin interaction, namely the hyperfine interaction with the nuclei in MoSe$_2$ or the exchange interaction with the magnetic ions of the EuS film. From the magnetic field angular dependence of the spin polarization we evaluate the anisotropy of the intervalley electron $g$-factor and the spin relaxation time. The non-zero in-plane $g$-factor $|g_x|\approx 0.1$, the value of which is comparable to its dispersion, is attributed to randomly localized electrons in the MoSe$_2$ layer. 
\end{abstract}

\date{\today}
\maketitle

Spin related phenomena in two-dimensional (2D) van der Waals semiconductors such as transition metal dichalcogenides have attracted close attention due to the unique energy level structure with spin-valley locking, resulting from strong spin-orbit interaction. The direct band gap excitons with large binding energy and oscillator strength determine the excellent optical properties of these materials~\cite{Wang-Chernikov-Glazov,Arora-review}. Novel spin phenomena and their application in spin-based photonic devices are of particular interest~\cite{Jin-2018,Sierra-2021,Lyons-2022,Ren-2022}. By now, the energy level structure of neutral and charged excitons as well as their spin and valley dynamics in monolayers are well studied. In particular, excitons in Mo-based dichalcogenide monolayers possess well defined selection rules for optical transitions, short lifetimes in the order of several ps, and fast spin relaxation due to electron-hole exchange interaction~\cite{Glazov-2014}. These properties can be used for efficient spin-valley pumping of resident carriers with circularly polarized light~\cite{Hsu-2015,Robert-2021}. Furthermore, optical experiments demonstrated exceptionally long spin lifetimes for electrons in the range of hundreds of ns and even longer for holes~\cite{Yang-2015, Marie-2017, Korn-2017, Crooker-2021}. In Refs.~\onlinecite{Yang-2015, Marie-2017}, anisotropic spin relaxation of electrons in 2D monolayers was used to explain the spin depolarization of resident electrons in a weak transverse magnetic field ($\sim$~10~mT). 

The spin dynamics in low dimensional nanostructures is very sensitive to the anisotropy of magnetic interactions and spin relaxation of both excitons and resident carriers. These features can be detected via spin depolarization in a transverse magnetic field (Hanle effect) and polarization recovery in a longitudinal field (polarization recovery effect)~\cite{OO}. The anisotropy is most clearly manifested in the angular dependence of the polarization curves, i.e. the transition from the Hanle to the polarization recovery effect in an oblique magnetic field (anisotropic Hanle effect). For excitons, the anisotropic Hanle effect in WSe$_2$ monolayer~\cite{Potemsky-2016} and CdSe/ZnSe quantum dots~\cite{Ku-Ku-} were reported using this approach. For electrons, the anisotropy of their $g$-factor in GaAs/AlGaAs quantum wells was evaluated~\cite{Kalevich-1992}. However, the investigation of resident carrier spins in oblique magnetic fields in 2D van der Waals materials has remained unexplored.

In this work, we study the spin polarization of resident electrons in a hybrid monolayer MoSe$_2$/EuS structure under resonant optical pumping in an oblique external magnetic field at cryogenic temperatures. We reveal an anisotropic Hanle effect for resident electrons in MoSe$_2$/EuS, which is result of a substantial anisotropy of both the electron g-factor and spin relaxation time. The Hanle effect is significantly impacted by the static random fluctuations of an effective magnetic field, which emerges due to the hyperfine interaction of localized electrons with the nuclei of MoSe$_2$ or the exchange interaction of the same electrons with the spins of the Eu ions in EuS, having a strength of several mT. This leads to an extraordinary Hanle effect: First, we observe a depolarization width of only a few mT in a transverse magnetic field, i.e. a long spin relaxation time of the resident electrons, which is supported by time-resolved pump-probe measurements. Second, in longitudinal magnetic field an even larger polarization recovery with comparable width is detected. This implies that the random fluctuations of the effective magnetic field acting on the localized electrons plays an important role in their spin depolarization in zero magnetic field. From the dependence of the spin polarization on the magnitude and orientation of the external magnetic field we evaluate the anisotropy of the intervalley electron $g$-factor and spin relaxation time. We emphasize that in our study the EuS film is not ferromagnetic and therefore there is no giant Zeeman splitting due to a magnetic exchange field as reported in Refs.~\onlinecite{Wei-2016, Zhao-2017, Norden-2019}. Nevertheless, the simple experimental approach based on single laser beam technique in combination with narrow Hanle curves has potential for application of 2D electrons in MoSe$_2$ for magnetic sensing.

\begin{figure}
	\includegraphics[width = 0.8\columnwidth]{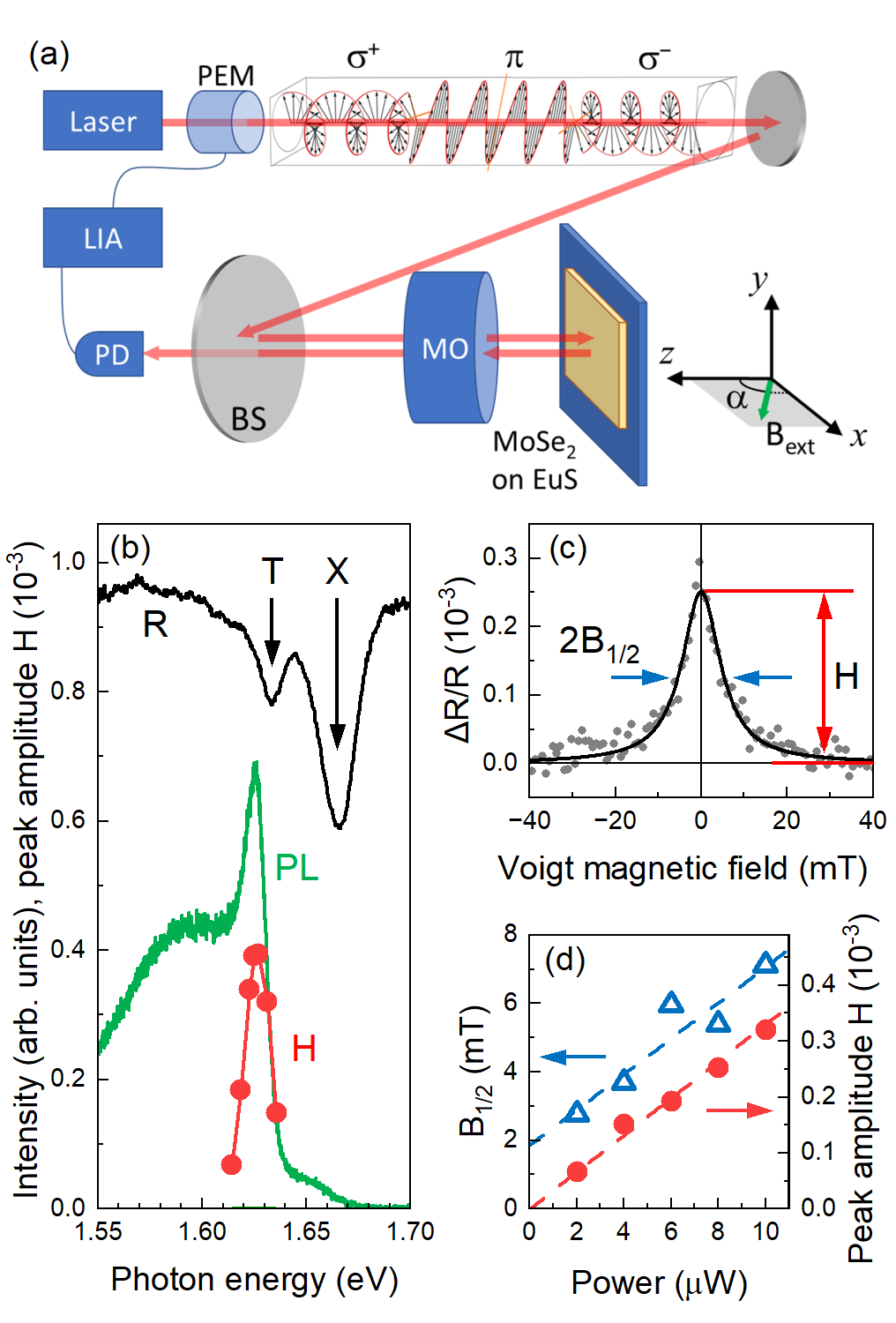}
	\caption{(a) Sketch of the experiment. MO -- microscope objective with 10x magnification, PEM -- photo-elastic modulator operating at 50~kHz, BS -- beamsplitter, PD -- photodiode, LIA -- lock-in amplifier. External magnetic field $\mathbf{B}_{\rm ext}$ is applied in the $xz$-plane at angle $\alpha$ with respect to the $z$-axis. (b) Photoluminescence (PL) and reflectivity (R) spectra measured at the temperature $T=5$~K.  (c) Hanle curve ($\alpha=90^\circ$, Voigt geometry) measured at the photon energy $\hbar\omega=1.631$~eV, excitation power $P=8~\mu$W, $T=2$~K. Solid curve is fit with Lorentz function with the half-width at half-maximum $B_{1/2} = 5.5$~mT and peak amplitude (height) $H = 2.5\times 10^{-4}$. Spectral dependence of the peak amplitude $H$ is shown in (b) by the circles. (d) Power dependence of $B_{1/2}$ (open triangles) and $H$ (solid circles) measured at $\hbar\omega = 1.631$~eV for $T=2$~K. In the limit of low power the magnitude of $B_{1/2}$ approaches 1.9~mT.}
	\label{Fig:sketch+Sum}
\end{figure}

The investigated sample comprises a single MoSe$_2$ layer on top of a 10~nm thick EuS film, deposited on a dielectric distributed Bragg reflector. The EuS film does not show ferromagnetic behavior but serves as source of electrons in the MoSe$_2$ layer with carrier density below $n_{\rm e} \approx 10^{12}$~cm$^{-2}$~\cite{Lyons-2022}. The experimental scheme is shown in Fig.~\ref{Fig:sketch+Sum}(a) and is based on single beam excitation and detection of the spin polarization of resident electrons as demonstrated earlier for $n$-doped CdTe quantum well structures~\cite{Saed-2019}. The resistive electromagnet is mounted on a rotating stage and the direction of the external magnetic field is applied in the $xz$-plane at a specific angle $\alpha$. The sample is mounted in a He-flow cryostat to keep it at low temperature ($T\sim 5$~K). Some measurements in Voigt ($\alpha=90^\circ$) and Faraday ($\alpha=0^\circ$) geometry were performed in a liquid-helium bath cryostat with a split-coil superconducting magnet, enabling higher magnetic fields and lower temperatures (down to $T =2$~K). The exciting laser beam is modulated between $\sigma^+$ and $\sigma^-$ circular polarization at 50~kHz using a photo-elastic modulator (PEM). The laser beam with normal incidence is focused into a spot with a diameter of about $4~\mu$m.  The reflected beam is detected with a photodiode (PD) and lock-in amplifier (LIA) at twice the PEM frequency.  Such detection scheme corresponds to differential reflectivity between circularly and linearly polarized excitation. For insightful results using pulsed excitation, the electron spin relaxation rate should be comparable to the laser repetition rate $f$, 80~MHz and 1~GHz in our experiment. Then spin polarization accumulates after excitation with multiple pulses and the measured signal $\Delta R$ is proportional to the resident electron spin density $S_z$ along the $z$-axis. 

The reflectivity (R) spectrum shows trion (T) and exciton (X) resonances located at photon energies of 1.633~eV and 1.665~eV, respectively, see Fig.~\ref{Fig:sketch+Sum}(b). The photoluminescence (PL) comprises the trion peak with a Stokes shift of about 7 meV and a broader emission band at lower energy $\sim 1.58$~eV. This band indicates the importance of localized states in the MoSe$_2$ monolayer. The Fermi level for resident electrons was estimated in Ref.~\onlinecite{Lyons-2022} to be about $E_{\rm F}\approx 5$~meV into the conduction band, which is smaller than the spin-orbit splitting of the conduction band states with opposite spins $\Omega_{\rm SO}=23$~meV~\cite{SO-splitting}. An examplary Hanle curve measured for $\alpha=90^\circ$ under pulsed excitation at $f=80$~MHz with photon energy $\hbar\omega=1.631$~eV using the light intensity $P=8~\mu$W is shown in Fig.~\ref{Fig:sketch+Sum}(c). The resident electron depolarization (decrease of $S_z$) is well described by a Lorentz curve with the half-width at half-maximum (HWHM) $B_{1/2}=5.5$~mT. 

The Hanle effect is observed in the vicinity of the trion resonance as follows from its spectral dependence in Fig.~\ref{Fig:sketch+Sum}(b). This can be explained by spin selective absorption in the probing process of the spin density $S_z$~\cite{Saed-2019}. The peak amplitude $H$ and $B_{\rm 1/2}$ grow linearly with $P$ as shown in Fig.~\ref{Fig:sketch+Sum}(d). The first insures light absorption  at the trion resonance in the linear regime, i.e. the density of excited trions is significantly smaller than $n_e$ and $\Delta R \propto S_z$, while the second relates to the reduction of the resident electron spin lifetime from trion excitation. In the limit of low powers the value of $B_{1/2}\approx1.9$~mT corresponds to the magnitude determined by the intrinsic spin relaxation time which is comparable with the values reported for electrons in MoS$_2$, WS$_2$, and WSe$_2$ monolayers~\cite{Yang-2015, Marie-2017}. We note that increase of the temperature up to $T=15$~K does not change the parameters of the Hanle curve.

\begin{figure}
	\includegraphics[width = 0.8\columnwidth]{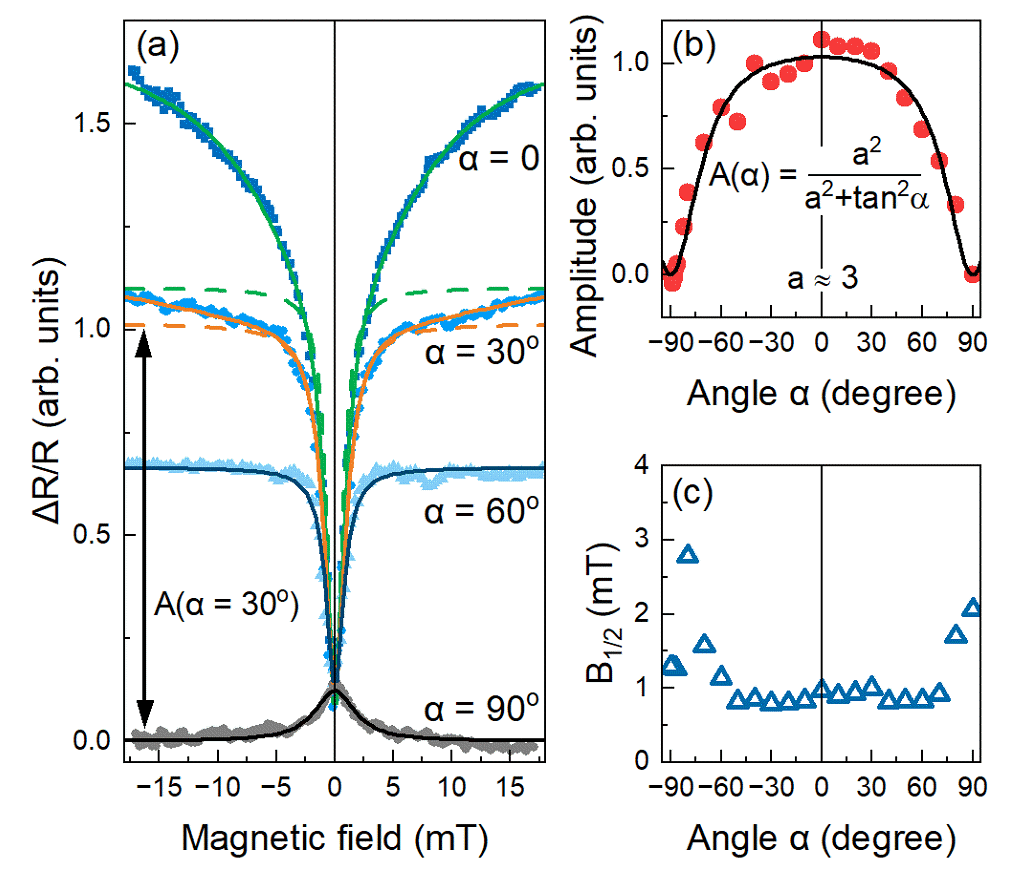}
	\caption{Hanle and polarization recovery effects. (a) Differential reflectivity $\Delta R/R \propto S_z$ as function of magnetic field for different angles $\alpha$. Voigt and Faraday geometry correspond to $\alpha=90^\circ$ and 0, respectively. (b) Angular dependence of amplitude $A(\alpha)$, as defined in (a), normalized to unity. (c) Angular dependence of half-width at half-maximum $B_{1/2}$. Temperature $T = 5$~K, photon energy $\hbar\omega = 1.6316$~eV, excitation power $P = 4~\mu$W, pulse repetition rate $f=1$~GHz.}
	\label{Fig:Hanle-vs-PRC}
\end{figure}

\begin{figure}
	\includegraphics[width = 0.6\columnwidth]{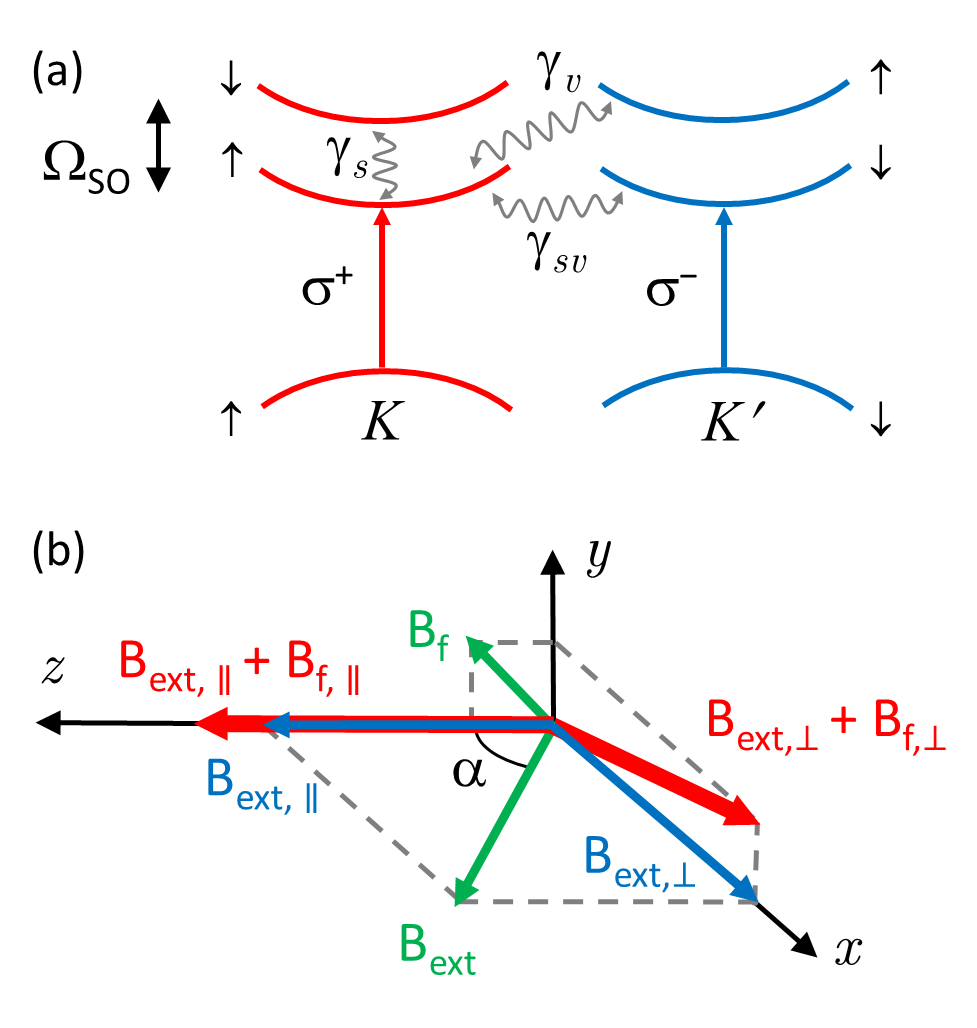}
	\caption{(a) Scheme of energy states in MoSe$_2$ monolayer. $\Omega_{\rm SO}$  -- magnitude of spin-orbit splitting, $\gamma_v$  -- spin conserving intervalley scattering rate, $\gamma_s$  -- spin relaxation rate within the same valley, $\gamma_{sv}$  -- intervalley spin relaxation rate. (b) External magnetic field $\mathbf{B}_{\rm ext}$ is applied in the $xz$-plane at angle $\alpha$ relative to the $z$-axis. $\mathbf{B}_{\rm f}$ is the random effective field. Red arrows indcicate longitudinal ($\|$) and transverse ($\perp$) components of the total magnetic field acting on the electrons.}
	\label{Fig:sketch2}
\end{figure}

The main result is presented in Fig.~\ref{Fig:Hanle-vs-PRC} where the variation of $S_z$ in magnetic field is shown for different angles $\alpha$. As discussed above, for a Voigt magnetic field ($\alpha=90^\circ$), the depolarization of the Hanle effect is observed. For Faraday ($\alpha=0$) and oblique magnetic fields we discover polarization recovery as manifested in an increase and saturation of $S_z$ with rising $B_{\rm ext}$. For angles $\alpha > 30^\circ$ the curves can be fitted by a single Lorentz curve with $B_{1/2}$ of only a few mT. For smaller angles ($0 < \alpha \leq 30^\circ$), an additional contribution to the polarization recovery is observed. This contribution has a larger HWHM of about 15 mT. 
In what follows we focus on the narrow dip in the magnetic field dependence and keep the broader polarization recovery out of consideration. In the Faraday geometry, strong polarization recovery is observed with the peak amplitude $H$ about 10 times larger as compared to the depolarization in the Hanle effect. Such behavior is surprising for two-dimensional semiconductors where a strong out-of-plane spin-orbit field $\Omega_{\rm SO}$ is present. Indeed, the spin-orbit field is parallel to the $z$ axis and does not cause relaxation of the $z$ component of the electron spin. Therefore, the magnetic field in Faraday geometry is not expected to influence the spin.

Spin depolarization of electrons in a weak transverse magnetic field was observed for AB$_2$ monolayers (A = Mo or W and B = Se or S) with $B_{1/2}\approx$ 10~mT in Refs.~\cite{Yang-2015, Marie-2017}. These results in combination with the absence of Larmor spin precession were explained by anisotropic spin relaxation~\cite{Oestereich-2004} with rates $\gamma_{x}=\gamma_{y}= \gamma_s+2\Gamma_v$ and $\gamma_z=\gamma_s$, where $\Gamma_v=\Omega_{\rm SO}^2/(4\gamma_v)$. Here,  $\gamma_v$ is the spin conserving intervalley scattering rate and $\gamma_s$ is the spin relaxation rate within the same valley. This mechanism requires $E_{\rm F} > \Omega_{\rm SO}$, which is not fulfilled in our case, and neglects intervalley spin relaxation $\gamma_{sv}$, see Fig.~\ref{Fig:sketch2}(a). 

Thus, suppression of spin relaxation and dephasing by an external magnetic field requires the presence of a different mechanism. Previously, an increase of the spin polarization of electrons localized on shallow donors in external magnetic fields of about 5 mT was observed in bulk GaAs~\cite{Dzhioev-2002}. The spin of a localized electron precesses in the random nuclear fields caused by hyperfine interaction. The longitudinal magnetic field along the initial spin with strength greater than the characteristic value of the random field $B_{\rm f} \sim$~5 mT, suppresses the precession of the $z$-component of the spin and restores its orientation (polarization recovery). On the other hand, a transverse magnetic field causes electron depolarization when it is stronger than $B_{\rm f}$ (Hanle effect). In our case, random magnetic fields in MoSe$_2$/EuS heterostructure can be created not only due to interaction of the electrons with the nuclei, but also due to exchange interaction with the magnetic atoms in EuS. In the following, we apply a model where the random fields are taken into account phenomenologically, without specifying their origin.

To analyze the Hanle and polarization recovery effects we use the Bloch equations that describe the evolution of the spin density of resident electrons $\mathbf{S}$~\cite{Glazov-08}:
\begin{eqnarray}
\label{Eq:main}
\frac{d\mathbf{S}}{dt} &=& \mathbf{G} + \mathbf{\Omega}\times\mathbf{S} - \mathbf{\gamma} \cdot \mathbf{S},
\end{eqnarray}
where $\mathbf{G} = (0, 0, G)$ is the spin pumping term.  Equation~\ref{Eq:main} is valid for low pumping rates $G/\gamma \ll n_e/2$. Spin pumping of the resident electrons is accomplished via excitation of trions with circularly polarized light. Due to spin relaxation in the trion state, repetitive excitation of spin polarized trions leads to dynamic polarization of the resident carriers~\cite{Henne-2006,Saed-2019}. Since $E_{\rm F}<\Omega_{\rm SO}$, we account only for resident electrons in the lowest energy states $K\uparrow$ and $K'\downarrow$ which are doubly degenerate at $B=0$. The intervalley spin relaxation is then determined by $\mathbf{\gamma}=\mathbf{\gamma}_{sv}$ which we assume to be anisotropic, i.e. $\gamma_\perp \equiv \gamma_x=\gamma_y$ and $\gamma_\| \equiv \gamma_z$. In order to explain the polarization recovery curves, we introduce the effective magnetic field $\mathbf B_{\rm f}$  fluctuating from one electron to an other and thereby leading to additional depolarization of the signal, even in zero external magnetic field. We also take into account the anisotropic electron g-factor $g_i$ which results in Larmor precession frequency components $\Omega_i = g_i \Omega_0 + bI_{i}$, where $\Omega_0=\mu_{\rm B} B_{\rm ext}$ with $\mu_{\rm B}$ being the Bohr magneton and $i=x,y,z$. The term $b \mathbf{I}$ accounts for the interaction with the random spin $\mathbf{I}$, $b$ characterizes the strength of the interaction. In our case this is a contact spin interaction, which requires electron localization~\cite{Avdeev&Smirnov}. 

The steady-state solution for $S_z$ in the absence of the effective field $\mathbf B_{\rm f}$ is given by 
\begin{eqnarray}
\label{Eq:SteadySz}
S_z=\frac{G}{\gamma_\|}\frac{\Omega_z^2 +\gamma^2_{\perp}}{\Omega_z^2+\dfrac{\gamma_\perp}{\gamma_{\|}} \Omega_x^2 + \gamma_{\perp}^2},
\end{eqnarray}
where $\Omega_x=g_x\Omega_0\sin{\alpha}$ and $\Omega_z=g_z\Omega_0\cos{\alpha}$. In our case $\mathbf{\Omega}$ is oriented in the $xz$-plane. In the general case, when the fluctuating fields are taken into account, the effective magnetic field acting on the electron spin is directed along an arbitrary direction, $\mathbf{B}=\mathbf{B_{\rm ext}}+\mathbf B_{\rm f}$, as shown in Fig.~\ref{Fig:sketch2}(b). However, due to the radial symmetry in the sample plane, it is enough to consider two components of $\Omega$, which is equivalent to rotation of the coordinate system around the $z$-axis. Therefore, we can use Eq.~\ref{Eq:SteadySz} and account for the fluctuating magnetic field by substituting $\Omega_z \rightarrow \Omega_{z}+bI_{z}$ and $\Omega_{x}^2 \rightarrow (\Omega_{x}+bI_x)^2 + (bI_y)^2$. 

The magnitude and direction of the fluctuating field are randomly distributed, which can be captured by a Gaussian distribution function~\cite{MER-2002}. However, in this case the magnitude of $S_z$ can be obtained only numerically. For this reason we use a simplified approach where averaging should result in vanishing of the terms linear to $\langle I_i \rangle =0$. The second order terms are replaced by the mean-square value $\langle I_i^2 \rangle = \frac{1}{3}I_0^2$~\cite{Dzhioev-1997}. In this case we obtain 
\begin{eqnarray}
\label{Eq:Final}
S_z=\frac{G}{\gamma_\|}\frac{\Omega_z^2+ \dfrac{1}{3}(bI_0)^2+\gamma^2_{\perp}}{\Omega_z^2 +\dfrac{\gamma_\perp}{\gamma_{\|}} \Omega_x^2 + \left[\dfrac{1}{3}+\dfrac{2\gamma_\perp}{3\gamma_{\|}}\right](bI_0)^2 + \gamma_{\perp}^2}.
\end{eqnarray}
Note that in the limit of large external magnetic fields, i.e. when $\Omega_0^2 \gg \gamma_{\perp}^2+(bI_0)^2$ the angular dependence of the relative spin density is given by the simple expression
\begin{eqnarray}
\label{Eq:Ratio}
A(\alpha)=\frac{S_z(\alpha)}{S_z(\alpha=0)}=\frac{\gamma_{\|}g_z^2\cos^2{\alpha}}{\gamma_{\|}g_z^2\cos^2{\alpha}+\gamma_\perp g_x^2 \sin^2{\alpha}}.
\end{eqnarray}
Using this expression we obtain the $a^2=\dfrac{\gamma_{\|}g_z^2}{\gamma_{\perp}g_x^2}=9$ using the fit of the amplitude $A(\alpha)=\dfrac{a^2}{a^2+\tan^2(\alpha)}$ as shown in Fig.~\ref{Fig:Hanle-vs-PRC}(b). This evaluation does not depend on the distribution of the fluctuating fields and was previously used to determine the anisotropy of the g-factor of localized electrons in a GaAs/AlGaAs quantum well structure~\cite{Kalevich-1992}.

Eq.~\ref{Eq:Final} allows one to explain the increase of $S_z$ (polarization recovery) in the Faraday geometry. Let us first compare the HWHM $\Omega_{0,1/2}=\mu_{\rm B}B_{1/2}$ observed in Faraday and Voigt geometry. According to Eq.~\ref{Eq:Final}
\begin{eqnarray}
\label{Eq:HWHM}
\Omega_{0,1/2}^{\rm F} = \frac{g_x}{g_z}\sqrt{\frac{\gamma_{\perp}}{\gamma_{\|}}}\Omega_{0,1/2}^{\rm V}=\frac{1}{g_z}\sqrt{\gamma_{\perp}^2+\left(\frac{1}{3}+\frac{2\gamma_\perp}{3\gamma_\|}\right)(bI_0)^2}
\nonumber,
\end{eqnarray}
where the superscripts "F" and "V" correspond to Faraday and Voigt, respectively. Taking into account that $\gamma_{\|}g_z^2 \approx 9 \gamma_{\perp}g_x^2$ we expect $\Omega_{0,1/2}^{\rm F} \approx \Omega_{0,1/2}^{\rm V}/3$, as indeed observed in the experiment shown in Fig.~\ref{Fig:Hanle-vs-PRC}(c) where the angular dependence of $B_{1/2}(\alpha)$ is shown. From these data we evaluate that the HWHM is about 2-3 times smaller in Faraday geometry compared to the Hanle half-width in Voigt geometry. Next, we estimate the upper value of $\gamma_\perp$ from the HWHM of the polarization recovery curve in Faraday geometry, $B_{1/2}^{\rm F}=1$~mT. The g-factor value for the lower energy conduction state of MoSe$_2$, $g_z = 3.68$, was evaluated from exciton reflectivity spectra in high magnetic fields~\cite{Koperski-2019}. Using this value, we obtain $\gamma_\perp < g_z \mu_B B_{1/2}^{\rm F} \approx 0.2~\mu$eV ($\hbar/\gamma_\perp  \gtrsim 3.3$~ns). Finally, the ratio between peak amplitudes in the Voigt and Faraday geometry, $C=H^{\rm F}/H^{\rm V}$ can be used to estimate the degree of anisotropic spin relaxation, as the following expression should hold $C \le 2\gamma_\perp / \gamma_\|$. As follows from Fig.~\ref{Fig:Hanle-vs-PRC}(a) $C\approx 10$ and therefore $\gamma_\perp \gtrsim 5 \gamma_{\|}$.  

\begin{figure}
	\includegraphics[width =0.8\columnwidth]{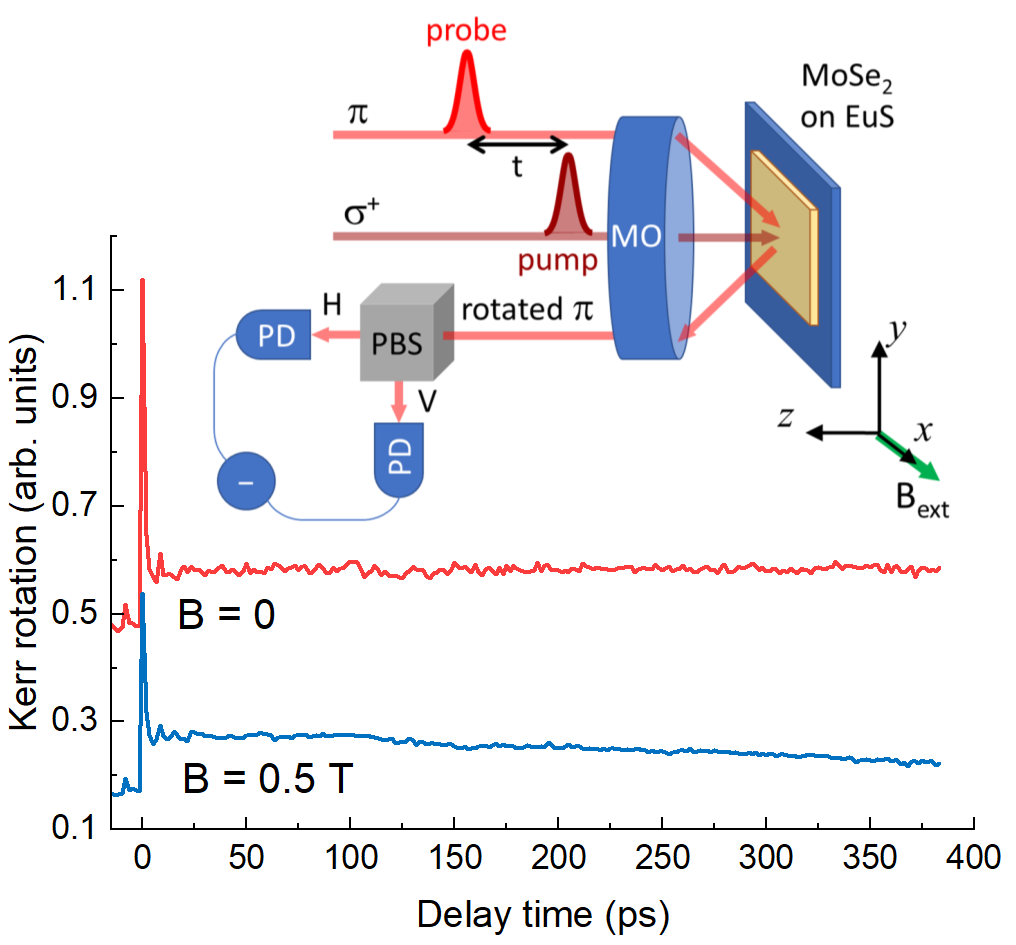}
	\caption{Time-resolved Kerr rotation in Voigt geometry for $B=0$ and $0.5$~T. The pump at photon energy $\hbar\omega_{\rm pump}=1.675$~eV is circularly polarized, the probe is linearly polarized at $\hbar\omega_{\rm probe} = 1.636$~eV. The rotation angle of the probe polarization is measured using balanced detection (see inset). PBS -- polarization beam splitter.}
	\label{Fig:Pump-Probe}
\end{figure}

Summarizing the above analysis, we are able to estimate the value of the transverse g-factor  $|g_x| \le \sqrt{\frac{2}{C}}\frac{g_z}{a} \approx 0.15$, i.e. $g_x \ll g_z$. In this case it is reasonable to assume that the dispersion of the transverse g-factor $\delta g_x$  due to inhomogeneities within the excitation spot is comparable with its magnitude. This can lead to the suppression of oscillations in spin transients. As shown in Fig.~\ref{Fig:Pump-Probe}, we indeed do not observe oscillatory behavior in a time-resolved pump-probe Kerr rotation experiment in Voigt geometry at $B=0.5$~T. In this case it is not necessary to fulfill the additional condition for the absence of Larmor prescesion $|g_x\Omega_0| < |\gamma_\perp - \gamma_\| |$, which would require an implausibly small $|g_x|<10^{-2}$ and a very long spin relaxation time $\hbar/\gamma_\| \approx 150~\mu$s~\cite{Oestereich-2004,Yang-2015,Marie-2017}. For a realistic spin relaxation time of $\hbar/\gamma_\| \approx 1~\mu$s, we estimate $|g_x|\approx\delta g_x \approx 0.1$. 

The analysis of the data clearly demonstrates that the spin dynamics in weak magnetic fields is determined by localized resident electrons. This argumentation is based on several reasons. In contrast to the strong spin-orbit splitting which is equivalent to an effective magnetic field of about 100~T acting on the electrons, a surprisingly small external magnetic field of only 1~mT leads already to the depolarization or polarization recovery of the resident electron spins. Spin conserving intervalley scattering is suppressed since the electrons do not occupy the higher energy states $K \downarrow$ and $K^\prime\uparrow$ and therefore cannot be responsible for the dynamical averaging of $\Omega_{\rm {SO}}$. The observation of comparable widths of the Hanle and polarization recovery curves is akin to the scenario of localized donor bound electrons in bulk GaAs where the spin relaxation is dictated by random nuclear fields~\cite{Dzhioev-2002}. We note, however, that the g-factor and spin relaxation are strongly anisotropic in 2D monolayers as compared to bulk semiconductors. Finally, the model requires a non-zero intervalley in-plane g-factor of the electrons. This is possible in case of their localization in the layer, which reduces the spatial symmetry and introduces mixing between the $K \uparrow$ and $K^\prime \uparrow$ as well as $K \downarrow$  and $K^\prime\downarrow$ states~\cite{communication-Glazov}. In our case, the breaking of symmetry is random and therefore the magnitude of the in-plane g-factor strongly fluctuates, which is manifested as spin dephasing without Larmor spin precession in the pump-probe experiment.  

To conclude we have demonstrated that localized electrons play an important role in the spin dynamics in 2D monolayers. In particular, we have shown that their spin relaxation in the studied MoSe$_2$/EuS structure is determined by random effective fields due to a contact spin interaction, namely the hyperfine interaction with the nuclei in MoSe$_2$ or the exchange interaction with the magnetic ions of EuS. Such localized electrons possess not only an anisotropic spin relaxation but also a non-zero out of plane g-factor due to mixing of the split-off bands with the same spins. The dispersion of the in-plane g-factor is comparable with its magnitude which is responsible for the absence of oscillatory behavior in the spin dynamics transients. The exact origin and magnitude of the random fields require further studies. The current work demonstrates that the implemented single laser beam technique is a powerful tool for spin studies in transition metal dichalcogenides. It opens new avenues for investigation of proximity effects in hybrid structures and magnetic sensor applications with 2D monolayers.

We are grateful to D.~S. Smirnov and M.~M. Glazov for useful discussions. The Dortmund and Sheffield groups acknowledge the EPSRC grant EP/S030751/1 support. This work was also partly supported by the Deutsche Forschungsgemeinschaft (project AK-40/11-1). I.V.K., O.S.K, V.L.K. acknowledge financial support from the Deutsche Forschungsgemeinschaft (project numbers 534406322,  529684269 and 524671439). O.M.H., D.J.G. and A.I.T. acknowledge support from the EPSRC grants EP/V006975/1 and EP/V026496/1.


\begin{thebibliography}{50}
%

\bibitem{Wang-Chernikov-Glazov}  G. Wang, A. Chernikov, M. M. Glazov, T. F. Heinz, X. Marie, T. Amand, and B. Urbaszek, {\it Excitons in atomically thin transition metal dichalcogenides}, Rev. Mod. Phys. {\bf 90}, 021001 (2018).

\bibitem{Arora-review} A. Arora, {\it Magneto-optics of layered two-dimensional semiconductors and heterostructures: Progress and prospects}, J. Appl. Phys. {\bf 129}, 120902 (2021).

\bibitem{Jin-2018} C. Jin, J. Kim, M. Iqbal Bakti Utama, E. C. Regan, H. Kleemann, H. Cai, Y. Shen, M. J. Shinner,
A. Sengupta, K. Watanabe, T. Taniguchi, S. Tongay, A. Zettl, and F. Wang, {\it Imaging of pure spin-valley diffusion
current in WS$_2$-WSe$_2$ heterostructures}, Science {\bf 360}, 893–896 (2018).

\bibitem{Sierra-2021} J. F. Sierra, J. Fabian, R. K. Kawakami, S. Roche, and S. O. Valenzuela, {\it Van der Waals heterostructures for spintronics and opto-spintronics}, Nat. Nanotechnol. {\bf 16}, 856 (2021).

\bibitem{Lyons-2022} T. P. Lyons, D. J. Gillard, C. Leblanc, J. Puebla, D. D. Solnyshkov, L. Klompmaker, I. A. Akimov, C. Louca, P. Muduli, A. Genco, M. Bayer, Y. Otani, G. Malpuech, 
 and A. I. Tartakovskii, {\it Giant effective Zeeman splitting in a monolayer semiconductor realized by spin-selective strong light–matter coupling}, Nat. Photon. {\bf 16}, 632–636 (2022). 

\bibitem{Ren-2022} L. Ren,  L. Lombez, C. Robert, D. Beret, D. Lagarde, B. Urbaszek, P. Renucci, T. Taniguchi, K. Watanabe,  S. A. Crooker, and X. Marie, {\it Optical detection of long electron spin transport lengths in a monolayer semiconductor}, Phys. Rev. Lett. {\bf 129}, 027402 (2022).

\bibitem{Glazov-2014} M. M. Glazov, T. Amand, X. Marie, D. Lagarde, L. Bouet, and B. Urbaszek, {\it Exciton fine structure and spin decoherence in monolayers of transition metal dichalcogenides}, Phys. Rev. B {\bf 89}, 201302 (2014).

\bibitem{Hsu-2015} W.-T. Hsu, Y.-L. Chen, C.-H. Chen, P.-S. Liu, T.-H. Hou, L.-J. Li, and W.-H. Chang, {\it Optically initialized robust valley-polarized holes in monolayer}, Nat. Commun. {\bf 6}, 8963 (2015).

\bibitem{Robert-2021} C. Robert, S. Park, F. Cadiz,  L. Lombez, L. Ren, H. Tornatzky, A. Rowe, D. Paget, F. Sirotti, M. Yang, D. Van Tuan, T. Taniguchi, B. Urbaszek, K. Watanabe, T. Amand, H. Dery, and X. Marie, {\it Spin/valley pumping of resident electrons in WSe$_2$ and WS$_2$ monolayers}, Nat. Commun. {\bf 12}, 5455 (2021).

\bibitem{Yang-2015} L. Yang, N. Sinitsyn, W. Chen, J. Yuan,  J. Zhang,  J. Lou, and S. A. Crooker, {\it Long-lived nanosecond spin relaxation and spin coherence of electrons in monolayer MoS$_2$ and WS$_2$}, Nature Phys. {\bf 11}, 830–834 (2015).

\bibitem{Marie-2017} P. Dey, L. Yang, C. Robert, G. Wang, B. Urbaszek, X. Marie, and S. A. Crooker, {\it Gate-Controlled Spin-Valley Locking of Resident Carriers in WSe$_2$ Monolayers}, Phys. Rev. Lett. {\bf 119}, 137401 (2017).

\bibitem{Korn-2017} M. Schwemmer, P. Nagler, A. Hanninger, C. Sch\"uller, and T. Korn, {\it Long-lived spin polarization in n-doped MoSe$_2$ monolayers}, Appl. Phys. Lett. {\bf 111}, 082404 (2017). 

\bibitem{Crooker-2021} Jing Li, M. Goryca, K. Yumigeta, H. Li, S. Tongay, and S. A. Crooker, {\it Valley relaxation of resident electrons and holes in a monolayer semiconductor: Dependence on carrier density and the role of substrate-induced disorder}, Phys. Rev. Materials {\bf 5}, 044001 (2021). 

\bibitem{OO} {\it Optical Orientation}, edited by F. Meier and B. P. Zakharchenya (Elsevier Science, North-Holland, Amsterdam, 1984).

\bibitem{Potemsky-2016} T. Smole\'nski, M. Goryca, M. Koperski, C. Faugeras, T. Kazimierczuk, A. Bogucki, K. Nogajewski, P. Kossacki, and M. Potemski, {\it Tuning valley polarization in a WSe$_2$ monolayer with a tiny magnetic field}, Phys. Rev. X {\bf 6}, 021024 (2016).

\bibitem{Ku-Ku-} Yu. G. Kusrayev, B. R. Namozov, I. V. Sedova, and S. V. Ivanov, {\it Optically induced spin polarization and g-factor anisotropy of holes in CdSe/ZnSe quantum dots}, Phys. Rev. B {\bf 76}, 153307 (2007).

\bibitem{Kalevich-1992} V. K. Kalevich and V. L. Korenev, {\it Anisotropy of the electron g-factor in GaAs/AlGaAs quantum wells}, JETP Lett. {\bf 56}, 257 (1992).

\bibitem{Wei-2016} P. Wei, S. Lee, F. Lemaitre, L. Pinel, D. Cutaia, W. Cha, F. Katmis, Y. Zhu, D. Heiman, J. Hone, J. S. Moodera, and C.-T. Chen, {\it Strong interfacial exchange field in the graphene/EuS heterostructure}, Nature Materials {\bf 15}, 711 (2016).

\bibitem{Zhao-2017} C. Zhao, T. Norden, P. Zhang, P Zhao, Y. Cheng, F. Sun, J.P. Parry, P. Taheri, J. Wang, Y. Yang, T. Scrace, K. Kang, S. Yang, G. Miao, R. Sabirianov, G. Kioseoglou, W. Huang, A. Petrou, and H. Zeng, {\it Enhanced valley splitting in monolayer WSe$_2$ due to magnetic exchange field},  Nature Nanothenology {\bf 12}, 757 (2017).

\bibitem{Norden-2019} T. Norden, C. Zhao, P. Zhang, R. Sabirianov, A. Petrou, and H. Zeng,  {\it Giant valley splitting in monolayer WS$_2$ by magnetic proximity effect}, Nature Communications {\bf 10}, 4163 (2019). 

\bibitem{Saed-2019} F. Saeed, M. Kuhnert, I. A. Akimov, V. L. Korenev, G. Karczewski, M. Wiater, T. Wojtowicz, A. Ali, A. S. Bhatti, D. R. Yakovlev, and M. Bayer, {\it Single-beam optical measurement of spin dynamics in CdTe/(Cd,Mg)Te quantum wells}, Phys. Rev. B {\bf 98}, 075308 (2018).

\bibitem{SO-splitting} A. Korm\'anyos, V. Z\'olyomi, N. D. Drummond, and G. Burkard, {\it Spin-orbit coupling, quantum dots, and qubits in monolayer transition metal dichalcogenides}, Phys. Rev. X {\bf 4}, 011034 (2014).

\bibitem{Oestereich-2004} S. D\"ohrmann, D. H\"agele, J. Rudolph, M. Bichler, D. Schuh, and M. Oestreich,
{\it Anomalous spin dephasing in (110) GaAs quantum wells: Anisotropy and intersubband effects}, Phys. Rev. Lett. {\bf 93}, 147405 (2004).

\bibitem{Dzhioev-2002} R. I. Dzhioev, V. L. Korenev, I. A. Merkulov, B. P. Zakharchenya, D. Gammon, Al. L. Efros, and D. S. Katzer, {\it Manipulation of the spin memory of electrons in $n$-GaAs}, Phys. Rev. Lett. {\bf 88}, 256801  (2002).

\bibitem{Glazov-08} G. V. Astakhov, M. M. Glazov, D. R. Yakovlev, E. A. Zhukov, W. Ossau, L. W. Molenkamp, and M. Bayer, {\it Time-resolved and continuous-wave optical spin pumping of semiconductor quantum wells}, Semicond. Sci. Technol. {\bf 23}, 114001 (2008).

\bibitem{Henne-2006} I. A. Akimov, D. H. Feng, and F. Henneberger, {\it Electron spin dynamics in a self-assembled semiconductor quantum dot: The limit of low magnetic fields}, Phys. Rev. Lett. {\bf 97}, 056602 (2006).

\bibitem{Avdeev&Smirnov} I. D. Avdeev and D. S.  Smirnov, {\it Hyperfine interaction in atomically thin transition metal dichalcogenides}, Nanoscale Adv. {\bf 1}, 2624 (2019).

\bibitem{MER-2002} I. A. Merkulov, Al. L. Efros, and M. Rosen, {\it Electron spin relaxation by nuclei in semiconductor quantum dots}, Phys. Rev. B {\bf{65}}, 205309 (2002).

\bibitem{Dzhioev-1997}R. I. Dzhioev, B. P. Zakharchenya, E. L. Ivchenko, V. L. Korenev, Yu. G. Kusraev, N. N. Ledentsov, V. M. Ustinov, A. E. Zhukov, and A. F. Tsatsul’nikov, {\it Fine structure of excitonic levels in quantum dots}, JETP Lett. {\bf 65}, 804 (1997).

\bibitem{Koperski-2019} M. Koperski, M. R. Molas, A. Arora, K. Nogajewski, M. Bartos, J. Wyzula, D. Vaclavkova, P. Kossacki, and M. Potemski, {\it Orbital, spin and valley contributions to Zeeman splitting of excitonic resonances in MoSe$_2$, WSe$_2$ and WS$_2$ Monolayers},  2D Mater. {\bf 6}, 015001 (2019).

\bibitem{communication-Glazov}  M. M. Glazov and D. S. Smirnov (private communication).

\end{thebibliography}
\end{document}